# A Novel Automated Classification and Segmentation for COVID-19 using 3D CT Scans

Shiyi Wang
National Heart and Lung Institute, Imperial College London
London, United Kingdom
s.wang22@imperial.ac.uk

Guang Yang
Cardiovascular Research Centre, Royal Brompton Hospital
National Heart and Lung Institute, Imperial College London
London, United Kingdom
g.yang@imperial.ac.uk

*Abstract*—Medical image classification and segmentation based on deep learning (DL) are emergency research topics for diagnosing variant viruses of the current COVID-19 situation. In COVID-19 computed tomography (CT) images of the lungs, ground glass turbidity is the most common finding that requires specialist diagnosis. Based on this situation, some researchers propose the relevant DL models which can replace professional diagnostic specialists in clinics when lacking expertise. However, although DL methods have a stunning performance in medical image processing, the limited datasets can be a challenge in developing the accuracy of diagnosis at the human level. In addition, deep learning algorithms face the challenge of classifying and segmenting medical images in three or even multiple dimensions and maintaining high accuracy rates. Consequently, with a guaranteed high level of accuracy, our model can classify the patients' CT images into three types: Normal, Pneumonia and COVID. Subsequently, two datasets are used for segmentation, one of the datasets even has only a limited amount of data (20 cases). Our system combined the classification model and the segmentation model together, a fully integrated diagnostic model was built on the basis of ResNet50 and 3D U-Net algorithm. By feeding with different datasets, the COVID image segmentation of the infected area will be carried out according to classification results. Our model achieves 94.52% accuracy in the classification of lung lesions by 3 types: COVID, Pneumonia and Normal. For 2 labels (ground truth, lung lesions) segmentation, the model gets 99.57% of accuracy, 0.2191 of train loss and 0.78±0.03 of MeanDice±Std, while the 4 labels (ground truth, left lung, right lung, lung lesions) segmentation achieves 98.89% of accuracy, 0.1132 of train loss and 0.83±0.13 of MeanDice±Std. For future medical use, embedding the model into the medical facilities might be an efficient way of assisting or substituting doctors with diagnoses, therefore, a broader range of the problem of variant viruses in the COVID-19 situation may also be successfully solved.

## I. Introduction

### A. Background

The coronavirus pandemic has had a devastating impact on global healthcare systems with a rapid outbreak in recent years and the continued emergence of variants of the virus. Subsequently, many experts have discovered that 3D computed tomography (CT) of the chest can be combined with deep learning to automatically detect and quantify lung lesions caused by the COVID virus. Radiologists have found that in the early stages, patients show multiple small patchy shadows and interstitial changes in the lungs, which are evident in the outer lung bands. After progression, patients with pneumonia are observed to have multiple hairy, infiltrative shadows in both lungs [1]. As these lung lesions can be observed in CT images, the images can be used by deep learning models to train and predict. For example, image classification of three categories: COVID virus, common pneumonia virus, and normal, as well as image segmentation of the infected region, left lung, and right lung. Ideally, these models can help doctors improve the efficiency and accuracy of diagnosis, and even identify areas of infection that are not discovered by doctors, which has profound implications for the monitoring and management of diseases caused by COVID.

### B. Related Work

A complete combined model incorporating both image classification and image segmentation can be used more efficiently in medical systems. In medical research, the classification of disease categories (e.g., brain tumor categories, lung pneumonia categories, etc.) and the segmentation of lesion regions have been developing relatively quickly in the field of deep learning. In previous studies, very few of them have been able to combine the two areas, for instance, to make the segmentation prediction on the branches of the classification results respectively. Also, as the accuracy of the prediction results is dependent on different datasets, exploring both different sizes of datasets and different labeled categories is extremely important for the accuracy of lung lesion prediction.

*1) Classification Models:* DL models have great performance in automated image classification on both X-Ray and CT images. A large number of DL methods are used for the classification of types related to the COVID pandemic such as COVID, Pneumonia, Normal, etc. AlexNet, Visual Geometry Group (VGG) network, GoogLeNet, DenseNet, XceptionNet, MobileNet, SqueezeNet, Inception-ResNet, CapsNet, NasNetmobile, ShuffleNet, EfficientNet and ResNet-50 have been widely used for image classification of X-Ray and CT images of COVID patients [2].

*2) Segmentation Models:* 3D U-Net is a rapidly developing and popular class of segmentation algorithms that can be applied to 3D images in recent years. At the same time, other algorithms such as VB-Net, FCN and SegNet also have good performance. MONAI Framework, an open-source AI framework from NVIDIA and King's College London in late 2019, is a free, community-supported, PyTorch-based framework for deep learning in medical imaging [3]. MONAI has a large number of composable and portable APIs for easy integration into existing workflows. It also offers multi-threading, which allows you to cache the converted data before training, and can greatly reduce the training time. On a convenience level, the DL network model can also be simply embedded into the MONAI framework.

## II. METHODOLOGY

### A. ResNet-50

The Resnet50 [4] network contains 49 convolutional layers and one fully connected layer. As shown in Fig. 1. below, the Resnet50 network structure can be divided into seven parts. The first part does not contain residual blocks and is mainly used for the computation of convolution, regularization, activation function and maximum pooling of the input. The second, third, fourth and fifth parts of the structure all contain residual blocks. The convolution blocks in the figure do not change the size of the residual blocks but are only used to change the dimension of the residual blocks [5]. In the Resnet50 network structure, the residual blocks all have three layers of convolution, so the network has a total of $1+3\times(3+4+6+3)=49$ convolutional layers, plus a final fully connected layer for a total of 50 layers, which is where the Resnet50 name comes from. The input to the network is 224 x 224 x 3, and after the first five convolutional layers, the output is 7 x 7 x 2048. The pooling layer converts this into a feature vector, and finally the classifier (fully connected layer) calculates this feature vector and outputs the category probabilities.

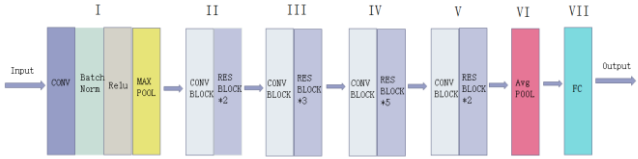

Fig. 1. The network structure of ResNet-50

### B. 3D U-Net

The core ideas of 3D U-Net are: First, to train on a sparsely labelled dataset and predict other unlabeled areas of this dataset. Second, train on multiple sparsely labeled datasets and then generalize to new data. Fig. 2. Shows the network structure of the 3D U-Net, which is similar to the 2D U-Net.

- There is an encoding path and a decoding path, each with 4 resolution levels.
- In the encoding path, each layer contains two $3\times3\times3$ convolutions (both followed by a ReLu layer) and there is a $2\times2\times2$ maximum pooling layer with a step size of 2 in each direction behind them.
- In the decoding path, each layer contains a $2\times2\times2$ deconvolution layer (up-conv) with a step size of 2, followed by two $3\times3\times3$ convolution layers, each followed by a RuLu layer.
- By shortcut, layers of the same resolution in the encoding path are passed to the decoding path, providing it with the original high-resolution features.
- The final layer is a $1\times1\times1$ convolutional layer, which reduces the number of output channels, the final number of output channels being the number of categories of labels [6].

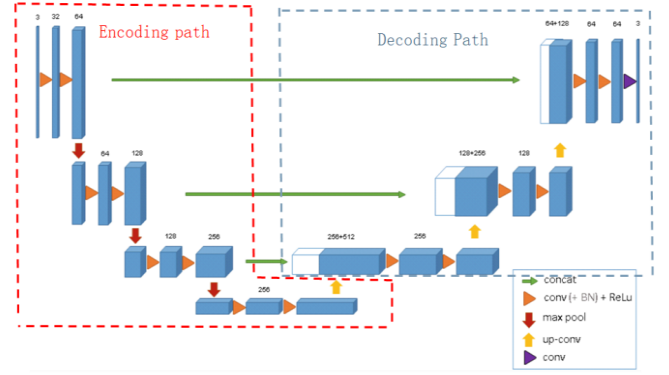

Fig. 2. The network structure of 3D U-Net

Compared to the 2D U-Net input, the 3D U-Net input is a stereo image (132 x 132 x 116) and is 3-channel. Also, batch normalization (BN) is added before ReLU, resulting in faster convergence. In addition, bottlenecks are avoided by doubling the number of channels before maximum pooling.

### C. MONAI

MONAI framework is built on PyTorch and aims to capture best practices in AI development for healthcare researchers with an immediate focus on medical imaging. MONAI also provides user-understandable error messages and an easy-to-program API interface. For example, our model builds deep learning training networks using the U-Net API in the MONAI framework. The CacheDataset function can provide multi-threaded processing, caching the transformed data before training, greatly reducing the unnecessary time wasted by loading data. The define event handler for training and validation of the model respectively. Set evaluator as a validator for the train handler. Create an object with images and labels for standard supervised training methods.

### D. Our Combined Model for Both Classification and Segmentation.

To solve the challenges of lung lesions in the COVID epidemic through deep learning. A complete medical system can be used for both classifications of lung pneumonia types and segmentation of lung lesions. As Fig. 3. indicates below, our combined model can realize both functions. First, for lung pneumonia classification, the original image will be pre-processed, some methods like position augmentation (includes scaling, cropping flipping padding, rotation, translation, affine transformation) and color augmentation (includes grayscale, contrast) is used in this step. It is also worth noting that data augmentation is only used on the training set and validation set, not the testing set. Then these images are fed to ResNet-50 for training, based on the performance, especially the accuracy, the best performance model is selected for classification prediction. The images being predicted may fall into one of the three categories, which are Pneumonia, COVID, and Normal. Since our experiment is about the segmentation of lung lesion regions, all images belonging to the "COVID" category are assembled based on the results of the classification and used as input to the segmentation algorithm.

For segmentation, the 3D image volumes are split into 2D slices, and the model uses MONAI's medical-specific transformations for image pre-processing. Then the concept of positive and negative maps mimics the human Electroencephalogram (EEG) model of positive and negative emotion recognition. The relevant annotation experts use interactive click annotation (sparse annotation) until the desired segmentation can be used. Positive guide maps are formed based on clicks and are used to guide (annotate) positive areas of interest, while additional negative guide maps represent non-lesioned areas (areas of disinterest). Then the data is as input for U-Net for training. After the training step, the best validation model is loaded for inference. The prediction of the segmentation has four labels which are ground truth, left lung, right lung and infected areas. Lastly, Colab visualizes results with scalars and continuous playback of 3D images (original and labeled images) by Tensorboard.

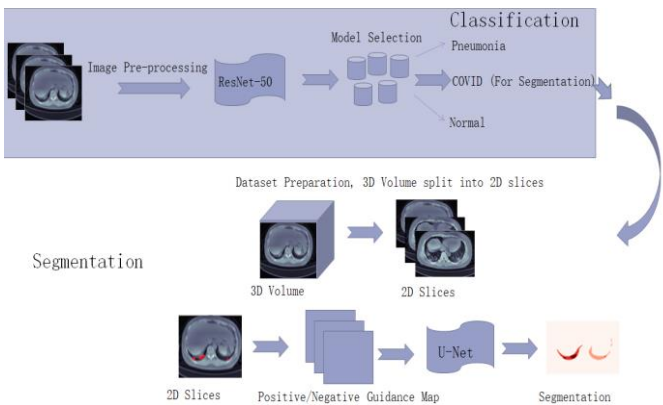

Fig. 3. Structure of our model

*E. Dataset and Tools*

1) Dataset "COVID-19&Normal&Pneumonia CT Images" [7] is used for classification. There are 2035 documents for COVID, 2119 for Normal, and 3390 for Pneumonia. 70 percent of the total dataset is used to train the model and the remaining 30 percent is used for validation.

Dataset " COVID-19 CT scans" [8] is used for four types of labels of segmentation (ground truth, left lung, right lung and infected area), there are 20 CT scans in total, 15 documents are for training, and 3 documents for validation and 2 documents for inference.

Dataset "COVID-19-20_v2" [9] is used for 2 types of labels of segmentation (ground truth and infected area), there are 199 documents in total, 160 documents for training, and 39 documents for validation.

2) Google Colab is used as an online computing tool to facilitate deep learning. Users can choose the RAM size and disk storage that suits their needs to train models using the Colab platform's GPUs. The code can be executed in the background with the browser closed. However, Colab also has the disadvantage that the supply used as a computing resource is not guaranteed and the link to the cloud will be automatically disconnected every 12 hours. In the meantime, the models are developed in the Phyton programming language to generate results with the PyTorch and Numpy libraries.

## III. EXPERIMENTS AND RESULTS

*A. 3 Types of Classification*

Different types of pneumonia can cause lung lesions of varying severity. Compared to common pneumonia, acute respiratory infections caused by COVID and its variants have a longer infection cycle, a much higher rate of transmission and mortality than common pneumonia viruses, and can even be severe enough to cause irreversible lung damage with corresponding sequelae. Therefore, it is extremely important to distinguish between the different symptoms of COVID and pneumonia, to treat the symptoms and prognosis of patients.

TABLE I. Performance of 3 types of classification

| Best Performance | Epochs | | |
|---|---|---|---|
| | *50* | *300* | *500* |
| Accuracy | 91.73% | 94.08% | 94.52% |

It is clear from TABLE I. that as the number of iterations (epochs) increases, so does the accuracy of the model prediction. However, the rate of increase from 300 to 500 epochs is slower than the rate of increase between 50 and 300 epochs. The performance of the model already achieves a good level at the 50th epoch with an accuracy of 91.73% and raises up to 94.52% until the 500 epochs.

*B. 2 Types of Segmentation*

In the two classes (infected area and ground truth) of lung lesion segmentation image experiments, in order for the model prediction to have high accuracy on 3D images as well, only a portion of 2D slices was input in the training phase after the stereo slices to generate dense stereo segments. It is also necessary for the data to be diverse, so a small number of 2D slices are taken from multiple samples to improve the generalization ability of the model. Fig. 4. has three parts: the original CT image (a), the input label (b), and the output (prediction) image (c). This experiment uses parameters which are: batch size =4, epochs=50, 100, 500, learning rate=0.0005, momentum=0.95. The same dataset and same epochs were used in the example given in the MONAI project from Github: the 2-label dataset and 500 epochs. Its model performance is lower than ours on the MeanDice metric, which is only 0.6904 ± 0.18 [10], while ours achieves 0.78±0.03. This is because it can be observed from the training curves that the model metrics start to converge at around 300 epochs, so adjusting the parameters accordingly can speed up the training. Our model adjusts the batch size from 2 to 4 and the learning rate from 0.0001 to 0.0005, and these two parameter changes result in improved performance.

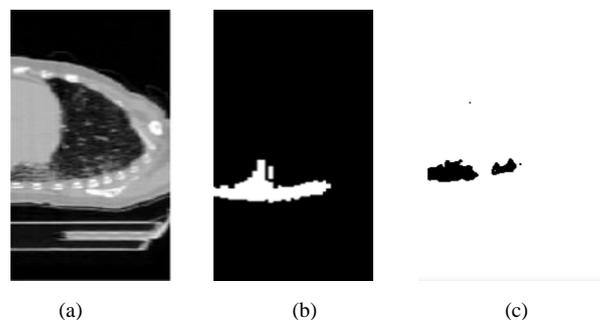

(a) (b) (c)
Fig. 4. The input and output of the segmentation model

According to Fig. 4. it is possible to observe in Fig. 4. (a). A part of the left lower side of the lung with an elongated faint ground glass shadow, which is the area of lesion caused by the infection of the lung by the COVID virus. Compared to the label Fig. 4. (b), Fig. 4. (c) can predict roughly the shape, but not for the details, e.g., the small triangles that protrude upwards. This may also be because it is not obvious in the original image (a) presented.

*C. Comparison of 4 Types and 2 Types of Segmentation*

Another dataset used for COVID lung lesion segmentation has 4 classes, which are ground truth, left lung, right lung and infected area. This experiment uses parameters which are: batch size =4, epochs=50, 100, 500, learning rate=0.0001, momentum=0.95.

TABLE II. Comparison of performance based on epochs

| Best Performance | Epochs | | | | | |
|---|---|---|---|---|---|---|
| | 50 | | 100 | | 500 | |
| | 2 Types | 4 Types | 2 Types | 4 Types | 2 Types | 4 Types |
| Train Loss | 0.3487 | 1.1195 | 0.2266 | 0.8988 | 0.2191 | 0.1132 |
| Accuracy | 99.14% | 95.93% | 99.28% | 96.34% | 99.57% | 98.89% |
| MeanDice ± Std | 0.68±0.04 | 0.50±0.05 | 0.72±0.04 | 0.59±0.07 | 0.78±0.03 | 0.83±0.13 |

For comparison, our experiment chooses three performances for 2 types and 4 types of segmentation on different epochs. (TABLE II.)

Train loss: The smaller the value of the loss function, the better the performance of the model in some respects. Because the four classes of annotation have more complex features at the level of the training step compared to the two classes, the training loss shows a sharp decrease with increasing epochs, while on the contrary the train loss curve for the two classes of segmentation shows a smooth performance. The training loss of 2 labels segmentation is from 0.3487 to 0.2191, while the 4 labels segmentation achieves from 1.1195 to 0.1132. It can be shown that the classification of 4 classes is easy to confuse at the beginning when the number of epochs is small, but as the number of epochs increases, its performance improves rapidly.

Accuracy: They both show great performance which is all over 95%, the 2 types of segmentation can even achieve nearly 100% of accuracy. Based on this curve variation, it can be seen that even though the segmentation model of 4 labels has a more complex data structure and a limited amount of data, it can still perform similarly well as the 2 labels segmentation. The accuracy of 2 label segmentation is over 99% on epochs 50, 100 and 500, from 99.14% to 99.57%. Also, the accuracy of the 4 label segmentation performed well, it was over 95% on epochs 50, 100 and 500, from 95.93% to 98.89%. Therefore, it can be seen that on epochs 50, both models can already achieve good accuracy.

MeanDice: The Dice Similarity Coefficient (DSC) is a set, similarity-based measurement method that is usually used to calculate the similarity of two samples, with values ranging from 0 to 1. The best segmentation performance is 1, and the worst is 0. First, the MeanDice score of 4 labels is lower than 2 labels, then it (4 labels) increased rapidly from 0.50 to 0.83, but the increased speed of 2 labels (MeanDice) is slower than 4 labels, which is from 0.68 to 0.78.

The results of the 4 class segmentation of lung lesions can be inferred by the selected model (Fig. 5.).

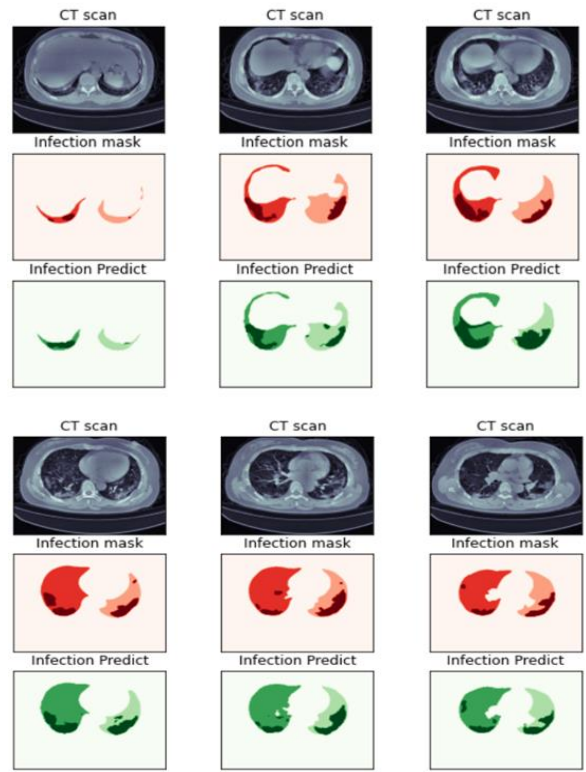

Fig. 5. The inference of 4 labels segmentation

The black images are slices from the original 3D CT scans. The red images show the left lung, right lung and infected areas (deep red color) respectively. The green ones are the predicted CT scans. By comparing the red and green images it can be seen that the model predictions are not as complete as the manually marked ones, but they are similar and the specific areas of infection are accurate and not missed. The model also has good performance in terms of measurement methods that were mentioned before (TABLE II.).

But we can also observe from the line graph (Fig. 6.) that the values of train loss (a) and MeanDice start to converge when the number of epochs is around 300. The training loss should be around 12k steps which are calculated as 300(epochs) * 40 (iterations)=12,000 when the batch size is 4. It is normal for metrics to fluctuate after the data has started to converge, as a small number of sparsely labeled datasets are used for prediction and the samples are diverse so that the model can generalize and the results become accurate.

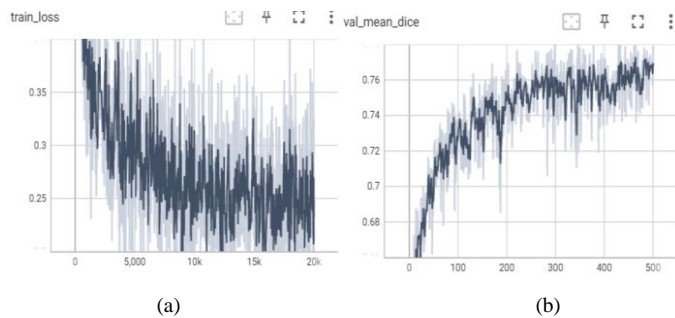

(a)          (b)

Fig. 6. Curves of train loss and MeanDice

It is worth mentioning that the datasets used for training in the experiments will not be used for validation, they are completely independent. Therefore, the experimental results and model performance are realistic and feasible.

## IV. Discussion and Future Work

There are several CT image features resulting from COVID pneumonia that can be annotated by relevant medical experts such as radiologists as input to the deep learning model. For example, CT imaging features of the lungs in patients with COVID may show: Lamellar and nodular ground glass shadows. Fine lattice-like shadows due to intralobular septal thickening. Multiple nodules in both lungs, etc. These features can be annotated and used for the diagnosis of COVID pneumonia. This, coupled with the efficiency and speed of the models trained by deep learning, can help doctors to better diagnose, manage and prognosticate the disease.

From the experiment, it can be claimed that the bigger epochs will lead to better performance. In terms of the size of the dataset, our 4-category label model, with only 20 images, outperforms the 2-category label model with 199 images in the final prediction. In addition, the datasets used in the experiment are from different COVID patients and contain different age groups, industries and genders. So the samples are more diverse and the predictions are more accurate without overfitting the model. Based on the inference of 4 label segments, our model can have a similar performance as human experts. When comparing the inference and true label masks, it can be observed that the inferred mask is extremely close to the true label, with the predicted lung lesion areas having approximately the same shape and location. It is also interesting to note that our model can also label small redundant areas. This is an interesting finding, as the redundant areas predicted by the deep learning model are not necessarily false positives, but maybe new lesions that have not been identified by the experts. This can be provided when the model is accurate and the annotation is accurate and diverse. This feature of deep learning models can therefore also help medical professionals to identify new lesions and thus improve the diagnosis, management and prognosis of COVID pneumonia.

However, there are also relevant restrictions. The biggest challenge should be the resources of the dataset. Medical annotation is a time-consuming and labor-intensive task that requires human experts to perform in order to ensure that the data is accurate and used for deep learning training. Otherwise, when annotations are unclear and error-ridden, the performance of the deep learning model degrades dramatically and the predictions become meaningless. Our model has the $0.78\pm0.03$ and $0.83\pm0.13$ MeanDice$\pm$std of 2 labels and 4 labels segmentation respectively, which can still be improved if the bigger datasets can be fed in theoretically. Generative Adversarial Networks (GAN) are used in some research. It is a good solution for generating the size of the dataset, but ensuring the availability of the generated labeled CT images remains a challenge for GAN. MONAI has also proposed a tool called MONAI label, which can be used in conjunction with MONAI's framework for training models to reduce the annotation time. In addition, a network called EcoNet, a scribble-based interactive segmentation method (no expert labels required), claims to have higher Dice scores with limited labeled data [11]. These methods need to be proven in practice, especially when applied to different model algorithms, so improving the performance of the models to make them more accurate and efficient to assist doctors in diagnosis is an important current research direction in the medical field. Embedding our models into facilities can help clinics that lack the medical expertise to perform COVID pneumonia diagnose. It can also be embedded into portable handheld CT scanners to facilitate patient self-diagnosis.